\documentclass[twocolumn,prb,superscriptaddress,longbibliography,amsmath,amssymb]{revtex4-1}
\usepackage{hyperref}
\usepackage{bm,upgreek}
\usepackage{graphicx}
\usepackage{color} 
\usepackage{setspace}
\usepackage{dcolumn}
\usepackage{subfigure}

\begin{document}
\title{Flat polarization-controlled cylindrical lens based on the Pancharatnam-Berry geometric phase}
\author{Bruno Piccirillo} 
\thanks{bruno.piccirillo@unina.it}
\affiliation{Dipartimento di Fisica ``E. Pancini'', Universit\`a di Napoli Federico II\\ Complesso Univ.\ Monte S. Angelo, via Cintia, 80126 Napoli, Italy}
\author{Michela Florinda Picardi}
\affiliation{Dipartimento di Fisica ``E. Pancini'', Universit\`a di Napoli Federico II\\ Complesso Univ.\ Monte S. Angelo, via Cintia, 80126 Napoli, Italy}
\affiliation{Department of Physics, King's College London, Strand, London, WC2R 2LS, United Kingdom}
\author{Lorenzo Marrucci}
\affiliation{Dipartimento di Fisica ``E. Pancini'', Universit\`a di Napoli Federico II\\ Complesso Univ.\ Monte S. Angelo, via Cintia, 80126 Napoli, Italy} 
\affiliation{CNR-ISASI, Via Campi Flegrei 34, Pozzuoli (NA) 80078, Italy} 
\author{Enrico Santamato}
\affiliation{Dipartimento di Fisica ``E. Pancini'', Universit\`a di Napoli Federico II\\ Complesso Univ.\ Monte S. Angelo, via Cintia, 80126 Napoli, Italy} 
\begin{abstract}
The working principle of ordinary refractive lenses can be explained in terms of the space-variant optical phase retardations they introduce, which reshape the optical wavefront curvature and hence affect the subsequent light propagation. These phases, in turn, are due to the varying optical path length seen by light at different transverse positions relative to the lens centre. A similar lensing behavior can however be obtained when the optical phases are introduced by an entirely different mechanism. Here, we consider the ``geometric phases'' that arise from the polarization transformations occurring in anisotropic optical media, named after Pancharatnam and Berry. The medium anisotropy axis is taken to be space-variant in the transverse plane and the resulting varying geometric phases give rise to the wavefront reshaping and lensing effect, which however depends also on the input polarization. We describe the realization and characterization of a cylindrical geometric-phase lens that is converging for a given input circular polarization state and diverging for the orthogonal one, which provides one of the simplest possible examples of optical element based on geometric phases. The demonstrated lens is flat and only few microns thick (not including the supporting substrates); moreover, its working wavelength can be tuned and the lensing can be switched on and off by the action of an external control electric field. Other kinds of lenses or more general phase elements inducing different wavefront distortions can be obtained by a similar approach. Besides their potential for optoelectronic technology, these devices offer good opportunities for introducing college-level students to an advanced topic of modern physics, such as the Berry phase, with the help of interesting optical demonstrations.\\[1 EM]
{\color{blue}
This is an author-created, un-copyedited version of an article published in European Journal of Physics, vol. 38, (2017) 034007 (15pp). IOP Publishing Ltd is not responsible for any errors or omissions in this version of the manuscript or any version derived from it. The Version of Record is available online at \href{https://doi.org/10.1088/1361-6404/aa5e11}{DOI: 10.1088/1361-6404/aa5e11}.}
\end{abstract}
\maketitle
\section{Introduction}
\label{intro}
In the last years, significant progress has been made in the field of ``structured light'', that is light featuring a relatively complex spatial structure. This structure may be in the intensity distribution (as in optical images) but also, and more interestingly, in the phase and polarization distributions of light. The phase distribution in the plane transverse to the main propagation direction defines in particular the wavefront structure, which in turn determines the subsequent propagation behavior of the light. Various families of structured optical beams with interesting properties have been demonstrated, including for example ``non-diffracting'' beams~\cite{durnin87} or helical-wavefront beams that carry well-defined values of the orbital angular momentum~\cite{piccirillo13a}. Structured light is also finding many applications: for example, in Ref.~\cite{vellekoop15}, it was shown that a suitably wavefront-shaped light beam can be focused inside even the most strongly scattering objects, overcoming the limitations imposed by light scattering to the depth at which optical imaging methods can retain their resolution and sensitivity. Some concepts and methods of the structured-light field have also been extended beyond the electromagnetic wave realm, as in the case of shape-invariant or helical electronic wavefunctions~\cite{uchida10,bloch13} or acoustic beams~\cite{zhang14}. The control of the wavefront structure can be nowadays based on a quite varied toolbox of methods, ranging from the simplest refractive lenses to the complex  ``transformation optics'' associated, for example, with metamaterial cloaking~\cite{kildishev08,service10}. Associated optical devices include refractive optical elements, such as for instance, axicon lenses~\cite{scott92}; diffractive optical elements, such as computer-generated holograms displayed on spatial light modulators~\cite{rosen95}; plasmonic or dielectric metasurfaces~\cite{bomzon01,yu11,kildishev13,yu14}; and finally Pancharatnam-Berry phase optical elements (PBOEs), which have recently stimulated a growing interest~\cite{pancharatnam56,bhandari97,marrucci11} and which are the subject of the present paper.

The working principle of ordinary refractive elements for wavefront shaping, such as standard lenses, is to introduce a space-variant phase retardation via a transversely-varying optical path length. This can be either from modulations of the thickness of the optical element, as in shaped glass lenses, or of the refractive index, as in gradient-index (GRIN) devices. As we will explain more in detail below, this phase can also be named ``dynamical'', as it is determined by the duration (in time or space) of the optical propagation process. In contrast, PBOEs exploit an additional phase retardation arising from polarization manipulations, which is independent of the propagation length. More in detail, in a PBOE, at each point in the transverse plane, the light polarization is made to change continuously along a given path on the Poincar\'e sphere by the action of a suitable local medium anisotropy. This sequence of polarization transformations in general gives rise to an extra optical phase retardation, in addition to the standard dynamical phase, that depends only on the geometry of the path of transformations on the Poincar\'e sphere, hence the name ``geometric phase''. This phase is also called Pancharatnam-Berry (PB) phase, from the names of the researchers who first investigated it~\cite{pancharatnam56,berry87}.

The first concept of a PBOE goes back to 1997, when Bhandari proposed the design of a geometrical-phase lens (GPL) consisting of a composite plate made of several concentric rings, each being 
a birefringent medium with half-wave retardation but having different orientations of the optic axis~\cite{bhandari97}, sandwiched between two standard (i.e. uniform) quarter-wave plates: a linearly-polarized optical beam passing through this device would acquire a geometric phase that varies in steps with the transverse distance from the center, giving rise, for suitable design of the structure, to a lensing effect. The lens would be focusing or defocusing, with the same focal length, depending on the input polarization, i.e. it is ``polarization-controlled''. In 2003, Hasman et al.~\cite{hasman03} experimentally demonstrated the first polarization-dependent GPL, in the form of a discretized PBOE manufactured with subwavelength gratings, operating in the mid-infrared. The first visible-domain GPL was realized with liquid crystals patterned by micro-rubbing in 2005 \cite{honma05}, although the authors of this work do not make explicit reference to the geometric PB phases to explain the device working principle. In 2006, Roux proposed the design of a continuous GPL for the visible domain that exploited the form birefringence of subwavelength grooves~\cite{roux06}. But only in 2015, Gao et al.~\cite{gao15} have reported the fabrication of liquid crystal GPLs exploiting the versatile polarization-holography alignment technique. At about the same time, liquid crystal GPLs based on a similar technology were also manufactured and used by our group for a proof-of-principle experimental demonstration of the concept of geometric-phase waveguides~\cite{slussarenko16}. 

In this paper, we illustrate the working principle, fabrication and characterization of a liquid-crystal cylindrical GPL fabricated via polarization-holography. The present GPL is also electrically tunable, i.e. its birefringent retardation can be controlled continuously via an external alternate-current (AC) voltage of few volts (at a frequency of few kHz), so that it can be operated with high efficiency across the whole visible and near-infrared spectrum. In addition, its optical action can be switched on and off by applying specific amplitudes of the same AC voltage. In this paper we chose to adopt the cylindrical lens geometry, as it involves only one-dimensional patterning and it is hence simpler to understand and characterize, but the same approach can be readily used to realize spherical lenses or more general optical phase elements.

The paper is organized as follows. In Sec.~\ref{sec:1}, we introduce the concept of PB optical phase. In Sec.~\ref{sec:2} we explain the working principle of PBOEs, which exploit the PB phase for controlling the optical wavefront, and then provide a mathematical description of the GPL. Next, in Sec.\ref{sec:3} we describe the experimental work, including the fabrication method (Subsec.~\ref{subsec:3_1}), the procedures adopted for the lens testing and characterization (polarization microscopy in Subsec.~\ref{subsec:3_2}, and a quantitative study of a Gaussian laser beam propagation and focusing after passing through the GPL (Subsec.~\ref{subsec:3_3}). In the concluding section, some brief remarks on the possible use of these devices for teaching purposes are given.

\section{The Pancharatnam-Berry phase}\label{sec:1}
In quantum mechanics, the adiabatic theorem states that a gradual change of the Hamiltonian from an initial form $H^i$ to a final form $H^f$ makes the system prepared initially in the (nondegenerate) $n^{th}$ eigenstate of $H^i$ slide continuously into the $n^{th}$ eigenstate of the Hamiltonian $H^f$, provided that no energy-level crossing occurs~\cite{griffiths94}. This is the basis for example for the well-known adiabatic approximation in molecular physics named as Born-Oppenheimer approximation. In 1984, Berry extended the adiabatic theorem by noting that, during the continuous transformation of the Hamiltonian, the eigenstates accumulate a global phase that can be given as the sum of two terms \cite{berry84}. The first, known as ``dynamical phase'', is the time integral of the (slowly varying) energy $E_n$ divided by the reduced Planck constant $\hbar=h/(2\pi)$, which is the obvious extension of the usual phase that is accumulated in time for the case of a static Hamiltonian and it clearly depends on the duration of the temporal evolution. The second term, in contrast, has no static analog and is determined only by the geometry of the path followed in the space of the parameters driving the slow evolution, while it is totally independent of the time duration: it is therefore named ``geometric phase'', or Berry phase, after its discoverer.

The Berry phase has been recognized as an important unifying concept across physics, not limited to quantum mechanics, with applications ranging from molecular spectroscopy to condensed matter, cold atoms, plasma physics, etc. For a comprehensive review about manifestations of Berry phase, see for instance Ref.~\cite{shapere89}. In the field of optics, the Berry phase has had a great impact, providing for example a unifying explanation for several recently understood ``spin-orbit optical phenomena'' \cite{cardano15,bliokh15}. In particular, the two most common manifestations of geometric phases in optics incude the Rytov-Vladimirskii-Berry phase (also known as ``spin redirection'' phase), arising when varying the wavevector direction of a light beam, and the PB phase, that is induced -- as mentioned in the introduction -- by a sequence of polarization manipulations occurring in anisotropic media \cite{pancharatnam,bhandari97}. In this paper, we are concerned only with the latter.

\begin{figure}
\includegraphics[scale=0.32]{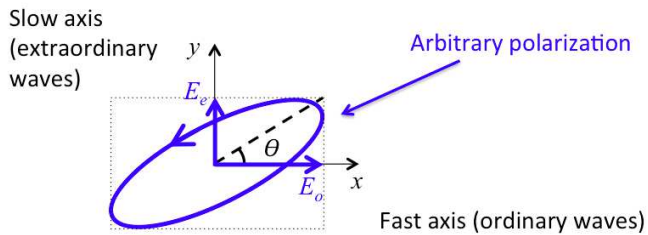} 
\caption{{\bf Elliptical polarization in birefringent medium and average index.} An arbitrary elliptically polarized wave traveling in a birefringent medium can be always decomposed in two orthogonal linearly oscillating fields, one along the direction of the fast optic axis (here named ordinary wave) and the other along the orthogonal slow optic axis (extraordinary wave). The ratio of the two projection amplitudes fixes the angle $\theta$, from which the average index $\bar{n}$ controlling the dynamical phase retardations can be defined.}
\label{fig1}
\end{figure}

The standard phase retardation accumulated by a light beam passing through a uniform medium is $\Delta\Phi = 2\pi n d/\lambda$, where $d$ is the medium thickness, $\lambda$ the vacuum wavelength and $n$ the medium refractive index (the product $nd$ is also named ``optical path length''). If the medium is birefringent, for any given propagation direction two distinct refractive indices can be defined, typically termed ordinary and extraordinary indices, here denoted $n_o$ and $n_e$. The \emph{dynamical phase} for an optical wave passing through such a birefringent (uniform) medium is then defined as the natural extension of the standard optical phase, i.e. $\gamma_d = 2\pi \bar{n} d/\lambda$, where $\bar{n}=\cos^2\theta n_o + \sin^2\theta n_e$ is the \emph{average refractive index}, $\theta$ being the angle formed by the input linear polarization direction with the ordinary (fast) axis in the transverse plane. More generally, if the input polarization is elliptical, $\theta$ is similarly defined by the projection angle of the input electric field onto the ordinary axis (see Fig.\ \ref{fig1}). In other words, the index average is weighted with the relative intensities $I_o \propto \cos^2\theta, I_e\propto \sin^2\theta$ of the ordinary and extraordinary components of the wave, i.e. $\bar{n}=(I_o n_o + I_e n_e)/(I_o+I_e)$ . For example, when using the standard birefringent waveplates, such as half-wave or quarter-wave plates, $\theta$ is determined by the orientation of the waveplate axis with respect to the input polarization. In particular, in all cases in which the waveplate axis is oriented at 45$^\circ$ with respect to a linear input polarization or if the input light is circularly polarized (in the latter case, for any orientation of the waveplate), the average index reduces to the arithmetic mean $\bar{n}=(n_o + n_e)/2$. Represented the polarization states as points on the Poincar\'e sphere, all these cases correspond to rotating around an axis that is orthogonal to the direction defined by the initial point on the sphere, and hence they correspond to moving along a geodesic arc on the sphere (i.e. an arc of a great circle)~\footnote{A geodesic arc between two points on the sphere (or a general curved surface) is the minimum-length path connecting the points: it is a concept that generalizes the notion of a `straight line' over a sphere (or a general `curved space').}.

\begin{figure}
\centering
\includegraphics[scale=0.6]{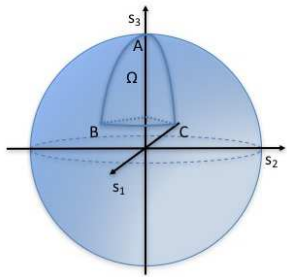}
\caption{{\bf Poincar\`e sphere and geodesic triangles.} The Poincar\`e sphere can be used to represent all possible polarization states of light. The cartesian coordinates in this representation are the reduced Stokes parameters $s_i=S_i/S_0$ with $i=1,2,3$, each ranging from $-1$ to $+1$. A closed curve on the surface formed by a sequence of polarization transformations determines a subtended solid angle $\Omega$ and the resulting geometric PB phase accumulated in the process. Particularly interesting are the curves formed by ``geodesic arcs'', because for such curves the dynamical phase is constant (for a fixed medium thickness). A geodesic arc between two points on the sphere is the minimum-length path connecting the points on the surface: it is a concept that generalizes the notion of a `straight line' for curved surfaces. In this figure, a geodesic `triangle' is depicted, connecting three points A, B and C with geodesic arcs and subtending a solid angle $\Omega$.}
\label{fig2}
\end{figure}

Now, as mentioned in the introduction, besides the dynamical phase defined above, birefringent media introduce an extra phase shift that is independent of the propagation length and is determined only by the geometry of the polarization-evolution path followed on the Poincar\'e sphere. More precisely, for any cyclic evolution starting and ending in the same polarization state, the extra phase over the cycle is given by the following simple expression: $\gamma_g = -\Omega/2$, where $\Omega$ is the solid angle subtended by the closed path (positive if the cycle is counterclockwise when seen from outside, negative if clockwise). This extra phase is just the PB geometric phase~\cite{pancharatnam,berry87}. For example, a sequence of three waveplates may bring the input polarization from a starting polarization $A$ to $B$, then to $C$ and finally back to $A$ along three geodesic arcs on the sphere. The PB phase $\gamma_g$ is then proportional to the solid angle $\Omega(A,B,C)$ of the ``geodesic triangle'' $ABC$ on the Poincar\'e sphere (see Fig.~\ref{fig2}). If the points $A, B, C$ are displaced, for example by a suitable rotation of the waveplates, but the connecting lines remain arcs of great circles, the phase change will only be given by the variations of $\gamma_g$, because the dynamical phase will stay constant.

The same definition of PB phase can be also applied to the \emph{phase difference} between two distinct optical transformations sharing the same initial and final states, as this can obviously be traced back to the phase variation of a single cyclic transformation by reversing one of the two transformations. This leads us directly to what is probably the simplest example of a PB geometric phase that can be introduced and easily controlled experimentally. This is associated to the transformation from a given input circular polarization state (left or right) to the opposite one, as obtained by the action of a single half-wave plate whose birefringent optic axis orientation is specified by the angle $\alpha$ (measured with respect to a fixed reference axis in the transverse plane). We are interested in the phase obtained by this transformation for an arbitrary angle $\alpha$, relative to that obtained for a reference angle, say $\alpha_0=0$. This corresponds in turn to the phase difference between two distinct transformations leading from left to right circular polarization or vice versa. For a circularly polarized input, the half-wave plate generates a geodesic arc on the Poincar\'e sphere, corresponding to the ``meridian'' located at 45$^\circ+2\alpha$ (relative to the reference direction), so that the two transformations subtend a solid angle $\pm 4\alpha$ and hence have a PB phase difference given by $\gamma_g = \pm 2\alpha$, where the sign is determined by the input polarization (in $\gamma_g$, it is $+$ for left circular input and $-$ for right circular input, where we adopt the naming convention corresponding to the point of view of the receiver). Hence, the optical phase of the outgoing wave can be controlled by simply rotating the half-wave plate. In contrast, the dynamical phase is independent of $\alpha$, so it plays no role.

\section{Controlling the optical wavefront: geometric-phase lens}\label{sec:2}
\begin{figure}
\includegraphics[scale=0.6]{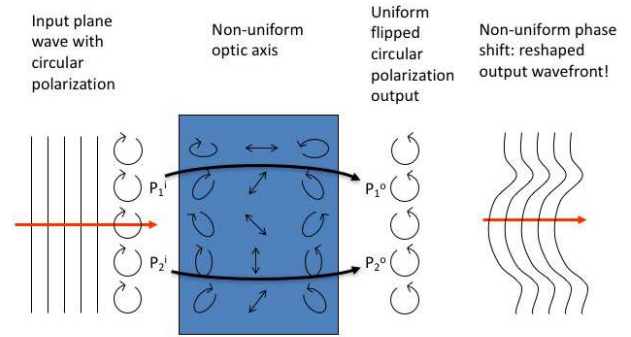}
\caption{{\bf Working principle of a PBOE.} A PBOE can be made as a birefringent medium with a uniform half-wave birefringent retardation but an optic axis that is space-variant in the transverse plane. An input circularly-polarized plane wave passing through the medium will be transformed into the opposite-handed circular polarization uniformly across the plate. However, the polarization transformations taking place in the medium are different at distinct points in the transverse plane and hence give rise to a space-variant transverse PB phase and a correspondingly reshaped output wavefront. For example, given any two transverse positions P$_1$ and P$_2$, the polarization evolution across the medium (black lines) is different and corresponds to two distinct meridians on the Poincar\'e sphere, sharing the initial and final points (i.e. the poles on the Poincar\'e sphere, corresponding to opposite circular polarizations). Hence, the two optical rays will acquire a relative PB phase difference given by half the solid angle subended by these two meridians.}
\label{fig3}
\end{figure}

The simplest possible design of a PBOE is a birefringent half-wave plate having a transversely space-variant optic axis, so that the PB phase discussed in the last example of the previous section varies in the transverse directions and results into a reshaped optical wavefront. In order to work properly and obtain only phase effects, the input light must be circularly polarized, and the output will be also circularly polarized with the opposite circular polarization (if different polarizations are instead used at input, one obtains a complex polarization transverse structure in the output). The resulting optical phase element has a uniform thickness, despite the arbitrarily large phase differences that can be induced across the plate. Another peculiar feature of this approach is that opposite handedness of the input circularly polarized light give rise to opposite phase retardations across the plate, and hence to ``conjugate'' wavefront outputs. The working principle of the device is illustrated pictorially in Fig.\ \ref{fig3}.

The standard materials used for making birefringent optics are crystals, whose optic axis cannot be patterned. In principle, one could obtain a ``segmented'' PBOE by simply attaching together several pieces of a crystalline half-wave plate, properly shaped and oriented, although this approach is clearly not very convenient. Modern approaches exploit different technologies to obtain patternable birefringent media, such as manufacturing subwavelength gratings or more complex metasurfaces, or using liquid crystalline materials. Liquid crystals are particularly well-suited for making PBOEs, as they are as birefringent as many ordinary crystals (actually, they typically have a fairly large birefringence), but their optic axis can be easily made non-uniform and controlled with patterned surface treatments or external fields.

In order to describe mathematically the action of the PBOE, we can use the Jones-matrix formalism~\cite{hecht}, in which the polarization state is described by two-component complex vectors giving the amplitudes of two prescribed orthogonal components of the total optical field. The choice of orthogonal polarization pair defines a basis in the Jones vector space. Anisotropic media are then simply described by two-by-two matrices, named Jones matrices. In particular, a birefringent medium seen in the ordinary-extraordinary linear polarization basis (as in Fig.\ \ref{fig1}) is described by the following diagonal Jones matrix (excluding the contribution of the dynamical phase, which is the same for all polarization components):
\begin{equation}
\bm{L}_0=\left(\begin{array}{cc} e^{-i\delta/2} & 0 \\ 0 & e^{i\delta/2} \end{array}\right).
\label{eq:HWPh}
\end{equation}
where $\delta=2\pi(n_e-n_o)d/\lambda$ is the birefringent retardation and we assume $n_e>n_o$ without loss of generality. For a half-wave retardation, $\delta=\pi$. The Jones matrix for an arbitrary $xy$ basis in the transverse plane can then be obtained from this by the action of the rotation matrix
\begin{equation}
\bm{R}(\theta)=\left(\begin{array}{cc} \cos\theta & -\sin\theta \\ \sin\theta & \cos\theta \end{array}\right),
\label{eq:rot}
\end{equation}
where $\theta$ represents the angle of the medium fast (ordinary) axis with respect to the $x$ axis. More precisely, this matrix will rotate the $x$ axis into the fast axis and the $y$ axis into the slow axis. Hence, the Jones matrix for a birefringent medium whose fast optic axis forms an angle $\theta$ with the $x$ axis is $\bm{L}_{\theta}=\bm{R}(-\theta)\cdot\bm{L}_0\cdot\bm{R}(\theta)$.

In a PBOE, the optic axis depends on the transverse coordinates, i.e. $\theta=\theta(x,y)$, so that this matrix $\bm{L}$ is also a function of $x,y$. We now make another basis transformation, from the linear-polarization basis to the circular-polarization (CP) basis. In particular, we introduce the complex unit vectors of the  left-/right- (L/R) circularly polarized states as follows: $\{\hat{L}=\left(\hat{x} + i \hat{y}\right)/\sqrt{2},\hat{R}=\left(\hat{x}- i \hat{y}\right)/\sqrt{2}\}$. The Jones matrix of the PBOE in the $\hat{L},\hat{R}$ basis is then given by
\begin{equation}
\bm{T}(x,y)=\cos{\frac{\delta}{2}}\left(\begin{array}{cc}1 & 0 \\0 & 1\end{array}\right)- i \sin{\frac{\delta}{2}}\left(\begin{array}{cc}0 & e^{-2 i \theta(x,y)} \\e^{2 i \theta(x,y)} & 0\end{array}\right).
\label{eq:PBOE}
\end{equation}
The PBOE is working properly -- i.e. it is properly ``tuned'' -- if $\delta=\pi$. In such case, the first term in the equation above vanishes and only the second term is present. A left-handed CP input wave with Jones vector $(1,0)$ in the CP basis, after passing through the PBOE is simply converted into the opposite CP state $(0,1)$ and multiplied by the space-variant phase factor $\exp{i2\theta(x,y)}$, which is the expected PB phase giving rise to the wavefront reshaping. If the input CP polarization is right-handed, the output will be left-handed and the phase factor becomes $\exp{-i2\theta(x,y)}$, i.e. the phase is sign-inverted and the conjugate wavefront is obtained.

When $\delta \neq \pi$, i.e. the PBOE is ``untuned'', only a fraction $\sin^2{\delta/2}$ of the input intensity suffers the phase change (which remains the same as for a tuned PBOE), while the remaining $\cos^2{\delta/2}$ is unaffected. The retardation $\delta$ depends on the medium thickness $d$, the wavelength $\lambda$ and the refractive indices $n_e$ and $n_o$. The last two, in particular, will depend on temperature and on the actual optic axis three-dimensional orientation, as for example its tilt with respect to the transverse plane $xy$. In liquid crystals, this tilt can be controlled electrically, and this yields a very convenient method for tuning $\delta$ and setting it to the desired value of $\delta=\pi$ for any given working wavelength~\cite{piccirillo10}.

\begin{figure}
\includegraphics[scale=0.57]{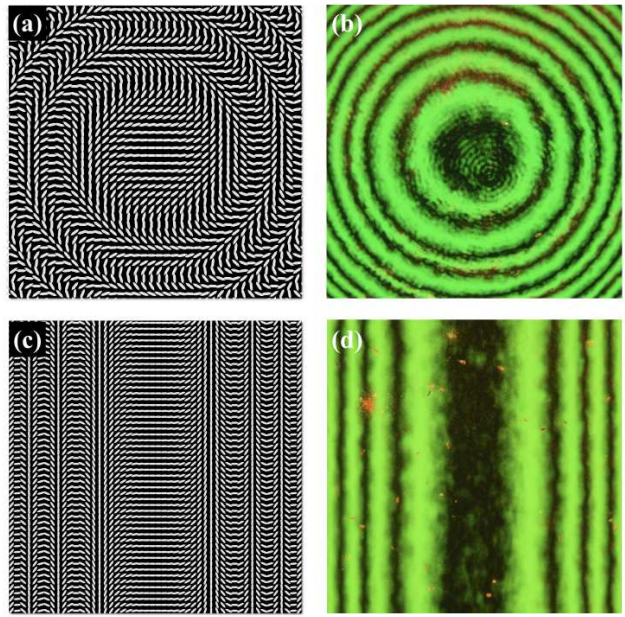} 
\caption{{\bf GPL optical structure.} Optic axis distribution of a spherical ({\bf a}) [cylindrical ({\bf c})] GPL between crossed polarizers and the corresponding microscopic images taken between crossed polarizers ({\bf b}) and ({\bf d})); dark fringes in the images correspond to locations where the optic axis is aligned either parallel or orthogonal to one of the polarizers.}
\label{fig4}
\end{figure}

Let us now specialize the discussion to a PBOE specifically designed to act as a focusing or defocusing lens, that is a GPL. This corresponds to saying that the PB phase must be quadratic in the transverse coordinates $x,y$ \cite{goodman,salehteich07}. Hence, we may generally set $\theta(x,y)=(\sigma/4) [(x / r_{0x})^2 +  (y / r_{0y})^2]$, where $r_{0x}$ and $r_{0y}$ are curvature radii along the two axes and the sign variable $\sigma=\pm 1$ defines the rotation direction of the optic axis (i.e. if the optic axis rotates clockwise or counter-clockwise when moving away from the origin). These parameters are constructive properties of the GPL. However, $\sigma$ can be sign-inverted by simply flipping the device by 180$^{\circ}$, or by reversing the propagation direction of the light through the device. For $r_{0x}=r_{0y}$, the GPL is circularly symmetrical and equivalent to a spherical lens, while for $r_{0x}=r_0$ and $r_{0y} \rightarrow \infty$, it becomes equivalent to a cylindrical lens with the axis along $y$. These two geometries are shown in Fig.\ \ref{fig4}.

\begin{figure}
\includegraphics[scale=0.58]{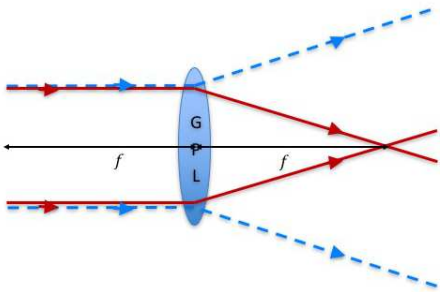}
\caption{{\bf Geometric-phase lens operation.} The geometric-phase lens (GPL) with circular polarization input behaves as a converging (continuous-line rays) or diverging (dashed-line rays) lens depending on the input polarization handedness. If the GPL is crossed in the opposite direction, the converging-diverging polarizations are swapped.}
\label{fig5}
\end{figure}

For an input CP wave, the phase retardation induced by this GPL device will be given by $\gamma_g(x,y)=\pm (\sigma/2) [(x / r_{0x})^2 +  (y / r_{0y})^2]$, where $\pm=+$ for an input left CP and $\pm=-$ for an input right CP. Therefore, the wavefront undergoes a change of curvature equivalent to that provided by a lens of focal distances $f_i = \pm \sigma \pi r_{0i}^2/\lambda$, with $i=x,y$ (see, for example, Refs.\ \cite{goodman} and \cite{salehteich07}). Hence, the sign of the focal distance of a GPL, determining if the lens acts as converging or diverging, will depend on both the side of the plate used as input and the handedness of the input CP wave. The polarization-controlled focusing/defocusing operation of the GPL is schematically illustrated in Fig.\ \ref{fig5}.

\section{Experiment}
\label{sec:3}
\subsection{Fabrication method}
\label{subsec:3_1}
Our GPLs were fabricated by using polarization holography in combination with the photoalignment of nematic liquid crystal cells~\cite{gao15,chigrinov08}. Our manufacturing procedure is aimed at realizing \O$1/2"$ lenses with prescribed FDs.

\begin{figure}
\includegraphics[scale=0.57]{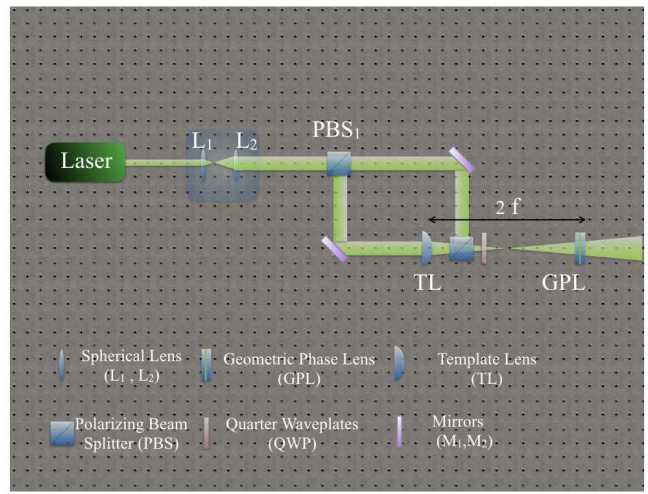} 
\caption{{\bf Schematic of the experimental apparatus for GPL photoalignment.} The light beam at $\lambda=532$~nm emitted by a frequency-doubled Nd:YVO$_4$ laser is magnified to half-inch diameter and injected into a polarizing Mach-Zehnder interferometer. The light beams propagating along the two arms are linearly polarized along orthogonal directions. A quarter waveplate is placed after the output polarizing beam splitter in order to turn the two superimposed beams into circularly polarized with opposite helicities . The template lens {\bf TL}  of (positive) focal lenth $f$ is placed in one of the two arms, while the photosensitive GPL cell (before filling it with liquid crystal) is located in the output beam after the quarterwave plate at approximately 2$f$ distance from the template lens.}
\label{fig6}
\end{figure}

Liquid crystal cells were initially realized assembling two parallel glass substrates separated by 6~$\mu$m with fixed spacers, which had been previously coated with a suitable photoaligning film~\cite{chigrinov08}. Before filling them with the liquid crystal, the empty cells were patterned by suitable exposure to an expanded laser beam (532~nm frequency-doubled Nd:YVO$_4$) with a prescribed inhomogeneous distribution of the linear polarization state corresponding to the desired pattern of the GPL optic axis. This polarization pattern is obtained in turn by using the output of a polarizing Mach-Zehnder interferometer (see Fig.~\ref{fig6}) having a standard glass template lens inserted into one of the arms. Let the phase difference between the Gaussian beam propagating in the reference arm and the lens-transformed Gaussian beam be denoted as $\psi(x,y)$. In polarization holography, the interference between the two circularly polarized beams having opposite helicities and phase difference $\psi(x,y)$, instead of an intensity modulation, gives rise to a space-variant linear polarization with a polarization direction oriented at angle $\psi(x,y)/2$ with respect to the reference axis. This distribution of polarization axes was then directly transferred into the optic axis of the GPL by exposing the photosensitive film of the coated glass slides to the interference field. After the exposure, the cell was filled with a nematic liquid crystal (commercial mixture E7, from Merck), by exploiting capillarity. The resulting geometric phase provided by the GPL is then $2 \theta(x,y)=\psi(x,y)$, that is the same as the phase difference between the two beams inside the interferometer.

Once the cell is assembled, the birefringent phase retardation $\delta$ of the GPLs can be controlled by applying a 10 kHz square-wave AC electric voltage with adjustable amplitude~\cite{piccirillo10}. The tuning condition $\delta=\pi$ (half-wave retardation) is obtained for $\approx 2.5$~V peak-peak. The GPL can also be optically ``switched off'' by applying the voltage giving full-wave retardation ($\approx 1.0$~V).

Let us call $f$ the focal distance of the template lens. The focal distance of the manufactured GPL needs not be exactly the same as that of the template lens, but it can be adjusted to a slightly different value $\bar{f}$ by placing the cell at a distance $f + \bar{f}$ from the template lens during the exposure stage. In fact, the radius of curvature of a Gaussian beam transformed by a lens of focal distance $f$ at a distance $f + \bar{f}$ from the lens is $\bar{f}$ up to terms of order $f^2 \lambda^2/(n^2 \pi^2 w_0^4)$, $w_0$ being the beam waist of the input beam on the lens. However, the dependence of the focal distance of a GPL on $1/\lambda$ requires more attention, since usually the cell is photoaligned using a light beam of wavelength $\lambda$ that is different from the actual wavelength $\lambda'$ of operation. To account for this mismatch, one can simply place the cell at a distance $f+\lambda' \bar{f}(\lambda')/\lambda$ from the template lens. In particular, to fabricate a GPL having focal distance $\bar{f}(\lambda')=f$, the distance of photoalignment is to be $f(1+\lambda'/\lambda)$. 

In our experiment, we used a template cylindrical lens of focal distance $f=150$~mm and fabricated a cylindrical GPL of nominal focal distance $\bar{f}=130$~mm at $\lambda=632.8$~nm.

\subsection{Characterization of GPL via polarization microscopy}
\label{subsec:3_2}
\begin{figure}
\includegraphics[scale=0.56]{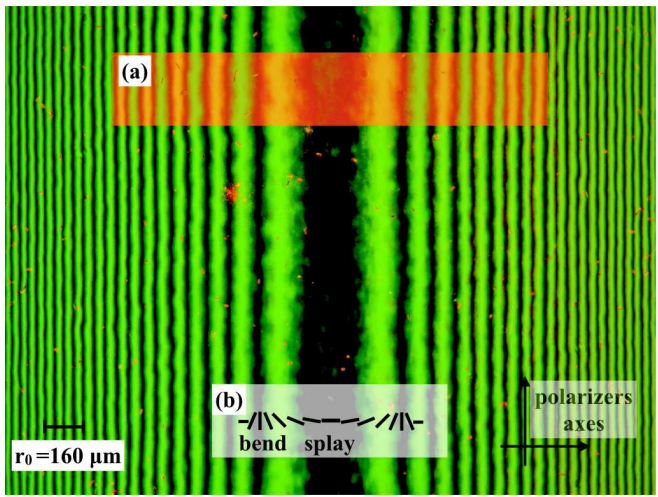}
\caption{{\bf Near field white-light image of the fabricated GPL between crossed polarizers}. The equivalent-curvature parameter $r_0 = 160~\mu$m of the GPL is reported. In inset ({\bf a}), the effect of the insertion of the $\lambda$-object before the analyzer is displayed. This image reveals that two adjacent maxima correspond to distinct values of the refractive indices and therefore correspond to orthogonal orientations of the optic axis. In inset ({\bf b}) the deduced distribution of the optic axis is also shown, noting the locations where splay and bend elastic reactions are induced in the liquid crystal.}  
\label{fig7}
\end{figure}

To fully characterize the fabricated GPL and check the quality of the product with respect to the design specifications, we measured the characteristic curvature parameter $r_0$. In Fig.~\ref{fig7}, we show the near field white-light image of our GPL between crossed polarizers, recorded by a digital camera connected to a polarizing microscope (Axsioskop, Zeiss) via a C-mount adapter.
The theoretical intensity profile along the $x$-direction perpendicular to the fringes can be obtained in the following way. Considering a linearly polarized input field, say along the $x$ axis, the transmitted optical field across the GPL can be computed with the help of the Jones-matrix formalism, using Eq.~(\ref{eq:PBOE}) with the parabolic expression of $\theta(x,y)$. The output field is then projected along the orthogonal direction $y$. The resulting intensity profile along the $x$-direction perpendicular to the fringes is described by the function
\begin{equation}
I(x)=I_0 \sin^2{\delta/2} \sin^2{(\frac{x^2}{r_0^2}+ 2 \theta_0)},
\end{equation}
where $\theta_0$ is an angle representing a rigid rotation of the optic axis distribution with respect to the axis of one of the crossed polarizers. The retardation $\delta$ includes the dependence on the wavelength, which turns out to be factorized with respect to the spatial modulation. This entails that the positions of intensity maxima $x_M$ (or minima $x_m$) are fixed. When the central fringe of the pattern is dark (central minimum), as is the case in Fig.~\ref{fig7}), $\theta_0 = 0$ and
\begin{equation}
x_M^2/r_0^2=(2 h +1)\pi/2,
\end{equation}
$h$ being an integer number. By fitting the position of the maxima $x_M$ (measured in pixel units, as obtained from the CCD camera image) as a function of the order of interference $h$ and accounting for the pixels-to-length conversion factor, we obtain the curvature radius $r_0=160.9 \pm 0.4$~$\mu$m. The error is derived from the data residual deviations from the best-fit. From this, the predicted focal distance of the GPL at $\lambda=632.8$~nm is $\bar{f}=128.45 \pm 0.07$~mm.

Measuring $r_0$ through this procedure has not only the merit of returning an accurate estimation of the expected focal length $f$, but also of testing the correctness of the lens quadratic phase profile, since this is related to the space-variation scale of the polarization fringes. Two adjacent maxima on the right (or on the left of the central minimum) correspond to two orthogonal orientations of the optic axis of the GPL, as can be easily inferred by the fringe pattern obtained inserting a $\lambda$-object in between the sample and the output polarizer\cite{bornwolf} (see inset (a) in Fig.~\ref{fig7}). The elastic reactions induced in the liquid crystals around adjacent maxima are different in nature -- splay or bend -- (see inset (b) in Fig.~\ref{fig7}). This observation may qualitatively explain the slight odd-even effect seen in the decreasing of the fringe width.

\subsection{Characterization of the GPL via beam propagation analysis}
\label{subsec:3_3}
\begin{figure}
\includegraphics[scale=0.56]{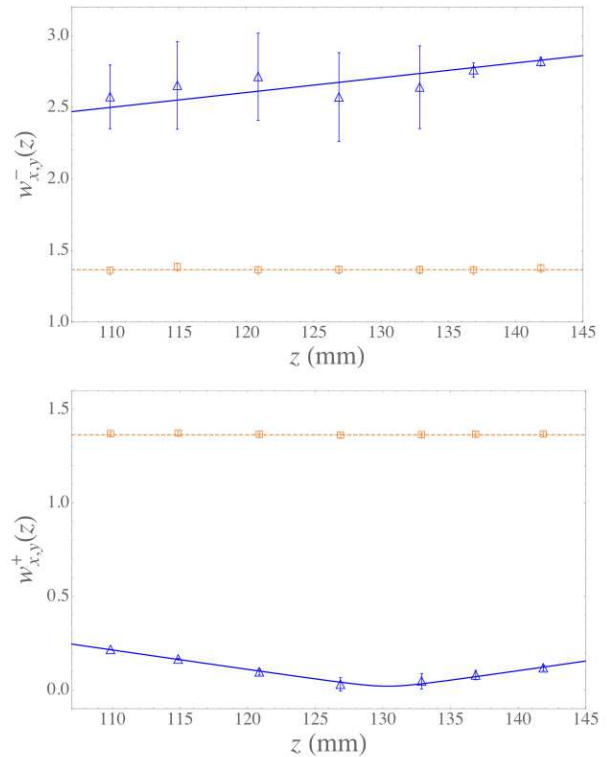} 
\caption{{\bf Beam radii beyond the GPL as functions of the distance $z$ from the lens}. ({\bf a}) Beam radii along $x$  ($\triangle$ for data points; continuous line for theoretical model) and along $y$ ($\Box$ for data points; dashed line for theoretical model based on Gaussian beam optics) for one of the two circular-polarization handedness at input. ({\bf b}) Beam radii along $x$  ($\triangle$ for data points; continuous line for the theoretical model) and along $y$ ($\Box$ for data points; dashed line for theoretical model) for the opposite handedness at input. Error bars are estimated from the fit residuals. It is evident that the radius along $y$ is always unaffected by the lens, while the radius along $x$ is decreasing to a minimum and the increasing again in the converging-lens case obtained for one circular-polarization handedness and monotonically increasing in the diverging lens case obtained for the opposite handedness.}
\label{fig8}
\end{figure}

As the next step, the optical lensing effect of the GPL was measured directly. A Gaussian beam (TEM$_{00}$ input mode) at $\lambda=632.8$~nm generated by a Helium-Neon laser was sent through the GPL and then imaged by a digital camera placed at various propagation distances after the lens. By analyzing the acquired images, we determined the corresponding beam radii $w_i^{\sigma}(z)$ with $i=x$ (focusing/defocusing direction) and $i=y$ (lens axis, with no focusing effect) for both left and right input circular polarizations. The input beam radius was also imaged and analyzed, yielding a beam-waist radius $w_0=1.478$~mm. All beam radii were obtained through Gaussian best-fits of the acquired transverse profile images (see Figs.~\ref{fig8} (a) and (b)). As expected, for each of the two input CP states, the beam radii vary as functions of the distance $z$ from the lens: the radius along the $y$ axis of the GPL remains unchanged for both polarization states (Figs.~\ref{fig8} (a) and (b), dashed lines); conversely, the radius along the $x$ axis perpendicular to the lens axis is increasing for a given input circular polarization (defocusing mode) (Fig.~\ref{fig8} (a), continuous line), while it exhibits a minimum for the orthogonal circular polarization (focusing mode) (Fig.~\ref{fig8} (b) continuous line). The beam radius $w_x^{\sigma}$ was also fitted as a function of the longitudinal coordinate $z$ according to the law
\begin{equation}
{w_x^{\sigma}}^2 = {w_{0 x}^{\sigma}}^2\left[ 1+ \left(\frac{z}{z_R}\right)^2\right],
\end{equation}
$z_R$ being the Rayleigh range of the gaussian beam. In the divergent case (Fig.~\ref{fig8} (a)), the beam waist (origin of $z$ axis) is virtually located behind the lens and the quadratic profile of $(w_x^-)^2$ as a function of $z$ cannot be experimentally observed around its minimum. Consequently, the best fit in this case is affected by larger uncertainties than in the case $(w_x^+)^2$, and this translates into larger best-fit-residuals assumed as errors on the experimental data points. From the position of this minimum we can determine the actual focal distance of the GPL, which is found to be $\bar{f}=130.3 \pm 0.3$~mm, very close to that predicted from the structural analysis. The small discrepancy between the two values is likely due to our small underestimation of the input beam divergence associated with the assumption we made that the input beam is perfectly Gaussian.

\section{Conclusions and final remarks}
In summary, after introducing the fundamental concept of the Pancharatnam-Berry optical phase, in this paper we explained the working principle of optical phase elements that exploit this phase, i.e. the PBOEs. We then restricted our attention to the specific case of geometric-phase lenses and illustrated a manufacturing method that makes use of polarization holography for patterning the optic axis of a liquid crystal cell using a standard glass lens as a template, leading to the preparation of a GPL that has approximately the same optical features. Following this approach, we prepared a cylindrical GPL and then characterized it by means of a direct structural analysis with polarized microscopy and by studying the induced focusing effects for different input circular polarizations.

We propose that GPLs such as that demonstrated in this paper might be used for teaching demonstrations or labs organized within a class of modern optics, or in support of an undergraduate-level introduction to the topic of Berry phase, within a course of modern physics or quantum mechanics. In the case of optics classes, we believe that the observation that a flat and extremely thin element can induce a very strong focusing, similar to that of a thick lens, can be very surprising, and it can trigger in the students useful discussions and reasoning about the role of optical phases in determining the subsequent propagation. The polarization dependence of the lens would also be surprising for most students, and it may give rise to unusual double-imaging effects when applied to natural (unpolarized) light, because a distinct image may be formed for each polarization component. When applied to illustrating Berry phases, these lenses may be used to vividly demonstrate the role that these phases may have in real-world effects.
\newpage
\noindent\textbf{Acknowledgments}
\noindent We thank Mr. Joseph Junca for useful discussions concerning the fabrication and testing of the GPL. This work was supported by the European Research Council (ERC), under grant no. 694683 (PHOSPhOR).
%
%
%
%
\end{document}